\documentclass[epj]{svjour}
\usepackage[utf8]{inputenc}
\usepackage{graphicx}
\usepackage[english]{babel}
\usepackage{fancyhdr}
\usepackage[]{xcolor}       
\usepackage{ulem} 
\usepackage{latexsym}
\usepackage{graphics}
\usepackage{multicol}
\defineshorthand{"~}{\babelhyphen{nobreak}}
\useshorthands{"}
\begin{document}

\title{Effects of temperature variations in high sensitivity Sagnac gyroscope}
\author{Andrea Basti\inst{1,2} \and Nicol\`o Beverini\inst{1,2} \and Filippo Bosi\inst{2} \and Giorgio Carelli\inst{1,2} \and Donatella Ciampini\inst{1,2}
\and Angela~D.V.~Di~Virgilio\inst{2} \and Francesco  Fuso\inst{1,2} \and Umberto Giacomelli\inst{1,2} \and Enrico Maccioni\inst{1,2}
\and Paolo Marsili\inst{1,2} \and Giuseppe Passeggio\inst{3} \and Alberto Porzio\inst{3,4} \and Andrea Simonelli\inst{2} \and Giuseppe Terreni\inst{2}}
%
%
\institute{ Dipartimento di Fisica Enrico Fermi, Universit\`a di Pisa,  Largo B.~Pontecorvo 3, Pisa, Italy
\and INFN Sez.~di~Pisa, Largo B. Pontecorvo~3, Pisa, Italy 
\and CNR-SPIN, Complesso Universitario di Monte Sant’Angelo, via Cintia, I-80126 Napoli, Italy
\and INFN Sez.~di~Napoli, Complesso Universitario di Monte Sant’Angelo, via Cintia, I-80126 Napoli, Italy}

%
\date{Received: date / Revised version: date}
%
\abstract{GINGERINO is one of the most sensitive Sagnac laser-gyroscope based on an heterolithic mechanical structure. 
It is a prototype for GINGER, the laser gyroscopes array proposed to reconstruct the Earth rotation vector
and in this way to measure General Relativity effects.
Many factors affect the final sensitivity of laser gyroscopes, in particular, when they 
are used in long term measurements, slow varying environmental parameters come into play.
To understand the role of different terms allows to design more effective mechanical as well as optical layouts,
while a proper model of the dynamics affecting long term (low frequency) 
signals would increase the effectiveness of the data analysis for improving the overall sensitivity. 
In this contribution we focus our concerns on the effects of room temperature and pressure aiming at further
improving mechanical design and long term stability of the apparatus.
Our data are compatible with a local orientation changes
of the Gran Sasso site below $\mu$rad as predicted by geodetic models. This value is, consistent with the requirements for GINGER 
and the installation of an high sensitivity Sagnac gyroscope oriented at the maximum signal, \textit{i.e.} along the Earth rotation axes.
\PACS{
} 
} 
\maketitle

\section{Introduction}
\label{intro}

 The family of inertial angular rotation sensors based on the Sagnac effect is rather large \cite{anderson1994}.
  They rely on the Sagnac effect that appears as a difference in the optical path between waves 
  (no matter if atomic or e.m. waves) propagating in opposite direction in a rotating closed loop. 
  This effect can be observed, like in the case of fibre gyroscopes \cite{Veliko2012}, as a phase difference 
  between the  two counter-propagating beams or, when the loop is a resonant cavity, as a frequency difference 
  between them.   In this last case, there are two possible strategies:
  to interrogate the cavities through external laser sources 
  (passive ring cavity, PRC \cite{Liu19}) or to observe the beat note between the radiation emitted in the two directions
  by a laser medium (active ring cavity, usually called ring-laser-gyro, RLG \cite{RSIUlli}).
  In principle a ring could operate both in active or in passive configuration, 
  and this is a very interesting feature for very high sensitivity measurements
  to get rid of systematic that affects differently active and passive devices.
  
  Eq. (\ref{eq:general}) reports the general relation connecting the Sagnac frequency $f_s $, the quantity measured for any RLG,
  and the modulus of the local angular rotation rate $\Omega$:
  \begin{equation}
      f_s = 4 \frac{A}{P \lambda} \Omega\cdot \cos(\theta)
      \label{eq:general}
  \end{equation}
  where $A$ is the area enclosed by the optical path, $P$ its perimeter, $\lambda$ the wavelength and $\theta$ the angle
  between the area vector and the local rotational axis.
  Since for a RLG rigidly connected to the ground, the Earth rotation velocity is by far the dominant component of $\Omega$,
  we can approximate $\theta$ with the angle between the area vector and the Earth rotational axis and define the RLG Scale Factor (SF) 
  as  $4A/(P \lambda) \cos(\theta)$.
    Any change in $f_s$ can be ascribed to different sources and, in particular for high sensitivity measurements,
  it is in principle very hard to discriminate between spurious rotations (affecting $\Omega$) 
  and changes in the scale factor, due to geometrical modifications and/or orientation changes,
  as they will produce the same effective change into $f_s$.
  For this reason, the details of the experimental apparatus matter in the final sensitivity, and even more in its long time response. 
   
\begin{figure*}
     \centering
     \includegraphics[width=\textwidth] {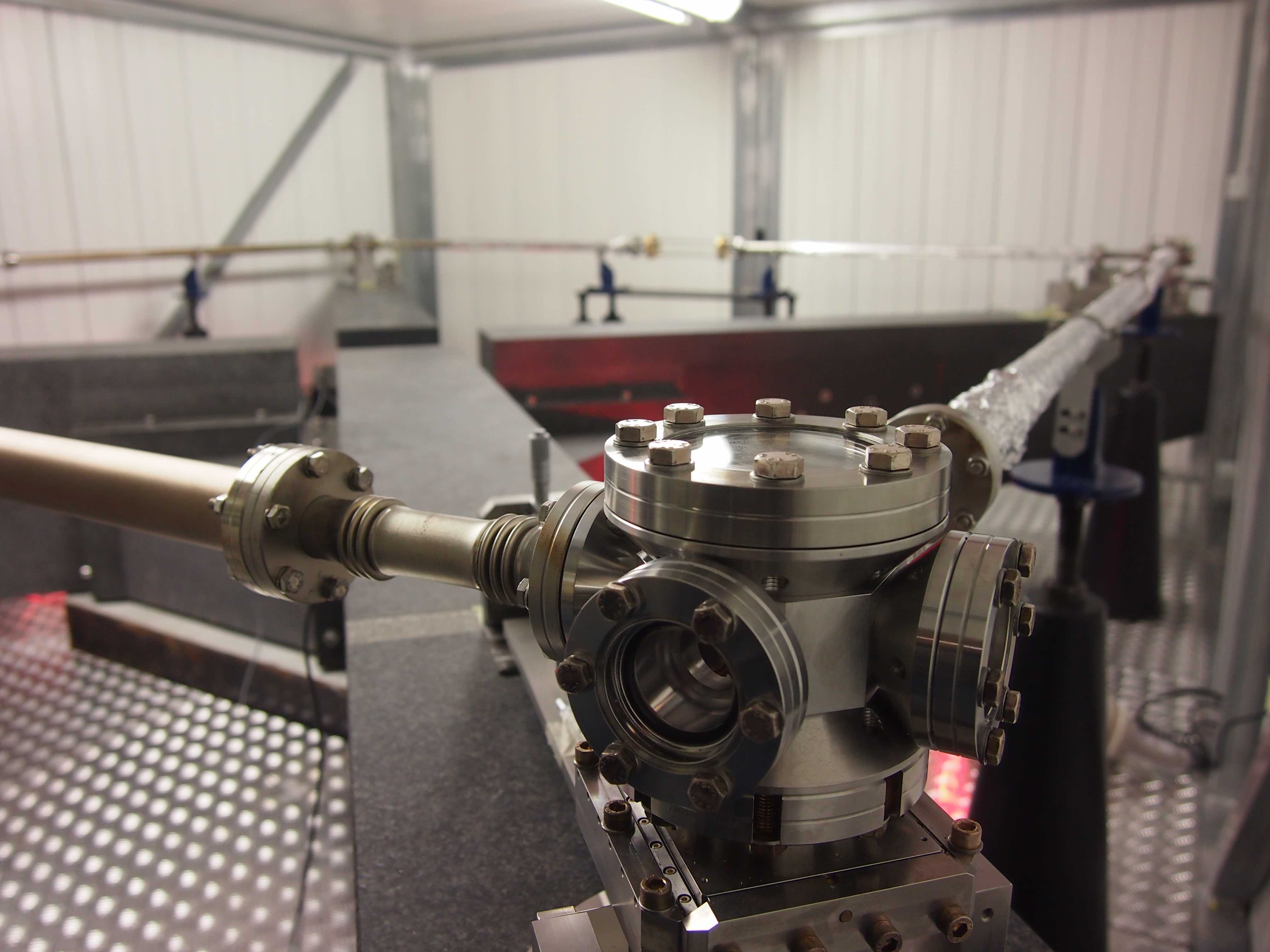} 
     \caption{GINGERINO at the time of assembling. In the foreground the mirror box and the pipes connecting the boxes can be seen,
     one of the Mitutoyo screw used to orientate the mirrors is visible beyond the box. It's a steel mechanical structure 
     attached to a cross shaped granite monument, which provide stability to the ring perimeter.}
     \label{fig:GINGERINO}
 \end{figure*}

Up to now, the most sensitive RLG was built with a so called monolithic design, a block of thermally stable material 
with optically contacted mirrors. 
This scheme is very expensive and, once installed, cannot be further enlarged or oriented at will \cite{RSIUlli}. 
Since RLG sensitivity increases with its dimension, a very large area, 833 m$^2$,
heterolithic (HL) RLG was built, whose mirrors were hold inside metallic boxes fixed to the floor \cite{Hurst2009}.
New HL apparatuses composed on a rigid monument supporting the mirrors metallic 
boxes have been developed in the last decade \cite{Belfi2012,90day,beverini_2020,igel2021}, (see fig.~\ref{fig:GINGERINO}).
Large frame high-performance HL passive optical gyroscopes have been also recently reported \cite{Martynov19,Zhang20,Liu19,Liu20,korth2016}.
 
 In HL structures, the whole light path is inside a single vacuum chamber where the mirror corner boxes are connected by pipes. 
 Mirror holders are also provided with mechanical and PZT driven tools to have a fine control on their position and orientation. 
 This control can be made active for geometry stabilization \cite{Santagata2015}.
 External disturbances can in principle produce spurious rotations and changes in the orientation of the area vector;
 moreover, couplings between the different mirrors are surely present, since there is a continuous mechanical connection, generating spurious signals
  GINGERINO is a highly sensitive  HL RLG prototype continuously running, unattended and without any active control.
 It has been assembled inside the underground INFN Gran Sasso laboratories (LNGS) 
in order to probe the site noise level and to study the potential sensitivity of HL RLG in the perspective
of the GINGER project, (Gyroscopes IN GEneral Relativity),
 which aims at measuring the Lense"~Thirring effect on the Earth with $1\%$ accuracy or even better \cite{prd2011,angela2017}.
Since the main goal of the GINGER project is to measure General Relativity and geodetic effects that produce DC or periodic signals, 
with periods ranging from few days to years, the investigation of the role of slow varying environmental perturbations,
such as temperature and pressure, becomes of great importance.
 The GINGER project is based on an array of RLG whose relative orientation can be chosen in order to optimally reconstruct the angular rotation.
 In particular, it is convenient to orientate one of the RLG at the maximum Sagnac frequency, 
 i.e. with the area vector parallel to the polar axis.
 In this conditions a variation of the absolute inclination of the RLG affects its scale factor only at the second order (see eq. 1),
 and it is possible to reconstruct the orientation with respect to the total angular rotation axis of the other RLGs \cite{angela2017,angelaFrontiers} to the micro-rad.
 This accuracy is required to measure the Lense-Thirring effect at the 1$\%$ level. In this contest, the stability of the underneath bedrock matters and should be further investigated.  
 
GINGERINO has clearly shown that HL mechanical structure can operate in the LNGS site with high sensitivity and long term stability
even if, in absence of an active control, mode jumps and split mode operation occasionally take place. These failures, however, affect the instrument duty cycle but not the sensitivity. 
It has been already proved that this perturbation typically  affects  $5\%$ of the data \cite{RSI2016}, thanks to the LNGS low environment noise.
However, it appears clearly that there is large room for improvements.  
 
The effects of temperature and pressure on the apparatus must be investigated in order to further improve the design in view of GINGER.
Clearly, data and findings should be handled with care since, as mentioned above, any change
 is totally equivalent to a  fluctuation of the angular velocity modulus.
 
 In the present paper the data analysis results are used to find bounds on the effect of temperature and pressure
 on the instrument sensitivity in order to assess the environmental constrains to be fulfilled by the GINGER design.
 
 It is well established that known geophysical signals induce local rotations and tilts that affect 
 in a well defined manner high sensitivity RLG. These signal are 
 for us a very useful for assessing the reliability of our instrument and check the effectiveness of data analysis.
 
 It is important to note that, while we draw specific conclusion in view of the GINGER design,
 the analysis and its procedure are more general and applies also to PRC.
 
 The paper is organised as follows. In Sect. \ref{Sect:Data} we review the elements of the quite complex data analysis we usually run on GINGERINO data.
 In particular, we focus on the role of temperature and its interplay with local tilts as disturbances on the Sagnac signal.
 Then, in Sect. \ref{Sect:Concl} conclusions are summarised.

\section{Data analysis}
\label{Sect:Data}
 
This paper aims at relating data coming from the long time observation of the environmental parameters with the Sagnac signal, 
We investigated the effects of temperature, pressure, air flow speed and local tilts whose probes are all co-located  with the RLG. 
The GINGERINO apparatus is contained inside a closed box made of thermal and acoustic shielding walls,
floor and roof, and it is composed by the vacuum chamber, 
whose corner are rigidly screwed to a granite structure attached to the bedrock through a central reinforced concrete support.
Temperature and pressure probes record the environmental data inside the box. 
A 2-channels tilt-meter is placed on the top of the granite monument to look for local tilts.
An anemometer is also installed in the tunnel outside the box to measure air-flow speed. All environmental data are sampled at 1 Hz.
The analysis has been applied to three time series (30 days in June 2018, 70 days in Autumn 2019, and 103 days in Winter 2020).
The RLG interference signal was elaborated following the technique described in details in Refs. \cite{Angela2019,Angela_epj2020,PRR2020,divirgilio2021}. 
There, we take into account the laser dynamic and reconstruct the true Sagnac frequency $f_s$, 
(in the following often expressed as the angular Sagnac frequency $\omega_s = 2\pi f_s$). 
So far, we have structured the analysis in order to recover from the data the global Earth rotation 
by taking into account laser dynamics and local disturbances coming from known geophysical signals.
Here, we aim at gain deeper inside the role of local environmental conditions, 
whose slow motion variations are indeed low frequency noise affecting the Sagnac signal. 
In particular, we look at correlation between environmental time series with the angular velocity resulting from the main data analysis.
In our analysis we consider $\omega_s$ as

\begin{equation}
	\omega_s = \omega_{geo} + \omega_{local}
\end{equation}

with $\omega_{geo}$ given by $2\pi\cdot SF\cdot\Omega_{geo}$, where $SF=A \cos \theta/(P \lambda)$ is the instrumental scale factor 
and $\Omega_{geo}$ is the global Earth rotation rate routinely 
measured and elaborated by the International Earth Rotation and Reference Systems Service (IERS). Moreovoer,
$\omega_{local}=2\pi\cdot SF\cdot\Omega_{local}$. In particular, $\Omega_{local}$ combines the rotational contribution of local geophysical origin as tides, ocean loading, etc.  
with instrumental spurious rotations that may be result from changes in the environmental operating condition. 
In principle, RLG data alone do not allow to discriminate between these two contributions.
It is then important to identify any kind of disturbances on the apparatus coming from the environment by using the data from environmental probes,
in order to improve the sensitivity and possibly the accuracy of future experimental apparatuses.

\subsection{Pressure and air--flow contribution}
\label{pressure}

We found no clear correlation of the gyroscope data with pressure and anemometer signals, notwithstanding the fact 
that GINGERINO box is not pressure isolated, 
by tight doors, from the tunnel as it is usual for these kind of high sensitivity instrumentation. 
As a matter of fact, while air flows variations are themselves negligible, pressure excursions are $\sim4 \%$. 
The lack of evidence of pressure fingerprint on the data does not rule out the possibility of an influence on the Sagnac signal. 
Surely, at low frequency, these variations does make sensible effect at the present stage. 
The pressure time line shows rather fast slopes most probably due to human activities that are not transferred to GINGERINO. 
However, pressure variations may be seen by the cave as a single mechanical structure with its own resonances,
so that a tight isolation is certainly wise for GINGER that will occupy a larger cave volume. 

\subsection{Tilts and local rotation}
\label{tilt}

\begin{figure*}
	\centering
	\includegraphics[width=\textwidth] {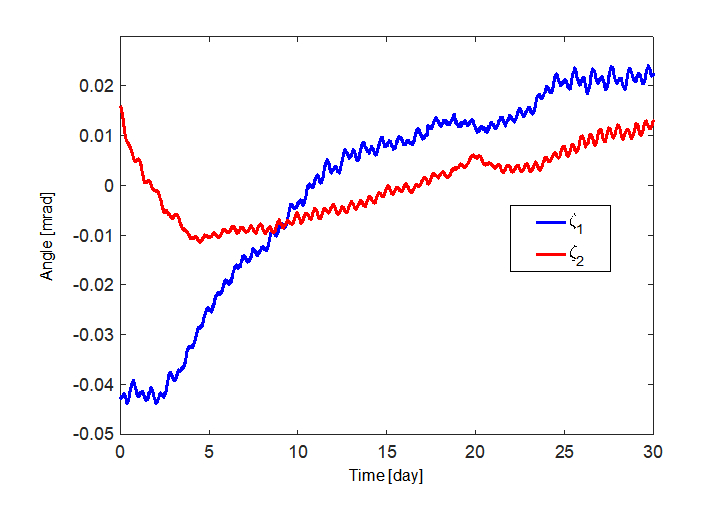}
	\caption{Time behaviour of the two tilt-meter channels during the analysed 30 days.}
	\label{fig:Tilt_s}
\end{figure*}

\begin{figure*}
    \centering  
    \includegraphics[width=\textwidth]  {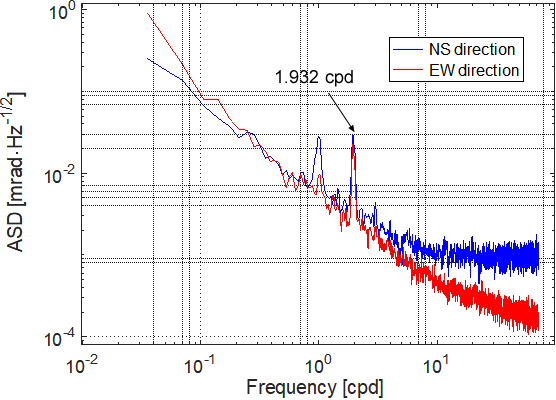} 
    \caption{Amplitude spectral density of the two tilt-meter channels during the first 103 days of 2020.}
    \label{fig:TiltASD}
\end{figure*}

A different finding came out for the two-axis tilt-meter and temperature data.
Any local rotation, due to known geophysical signals and/or instrumental and environmental noise results in changing the effective SF
that includes the orientation of the ring plane with respect to the actual rotation axis and the geometry of the cavity.
Preliminary we note that tilt-meter data give a clear signal of the Earth solid tides.
The tides show-up as tiny bi-daily oscillation in both time series of Fig. \ref{fig:Tilt_s}. 
The period of such oscillations can be better estimated by looking at the amplitude 
spectral density (ASD) of tilt-meter reading for the two axis (N--S and E--W respectively) shown in Fig.\ref{fig:TiltASD}. 
These plot refer to the data collected in the 103 days long run in 2020. Such a long observation period allows a much better spectral
resolution of low frequency signal such as solid tides.

Indeed, the ASD of tilt-meter data shows a peak at at 1.932 cpd (corresponding to a period of 12.42 hours),
that corresponds to the frequency of the main tide component. 
Before going into details of the interplay between temperature and tilts, we stress that our tilt meter signals are quite similar 
to the typical geophysical signal recorded elsewhere by other high sensitive instrument. For instance the records of the clinometer Marussi, 
installed in the cave of Bus de la Genziana (Friuli,  Italy) \cite{Devoti,devoti2019}, show that most of the tilts of the monument can be assumed to have geophysical origin. 
We will see that long time drift may be ascribed to temperature changes. We hereby stress that known geophysical signals play the role of a “test signal” 
for our apparatus giving us the possibility of evaluating its reliability in view of GINGER.
	
\subsection{Temperature Effects}
\label{temperature}

As above said the analysis evaluates $\Omega_{local}$, the total local disturbance, using these information 
and the environmental monitors to recover at our best the Earth rotation velocity.
In this contest, known geophysical signals  are of great importance since they can be used to increase the instrument accuracy and sensitivity.
Any GR effects is contained into the Earth rotation signal once it's cleared from any local perturbation, that's why
 the study of room temperature effects on the apparatus is a key point. 
The temperature variations are very slow, for this reason they have a great importance in  the frequency region where the GR terms should appear.

\begin{figure*}
    \centering
    \includegraphics[width=\textwidth] {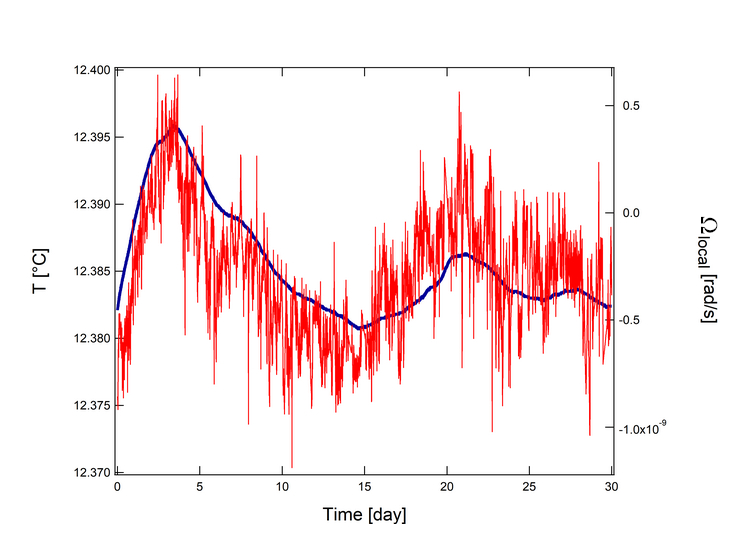}  
    \caption{Temperature in $^{\circ}$C (blue line), and $\Omega_{local}$, in rad/s (red line) from the 2018 30 days data set }
    \label{fig:Omega_T}
\end{figure*}

A typical temporal behavior of temperature and $\Omega_{local}$ is shown in Fig. \ref{fig:Omega_T} that compares
the time series obtained from the 30 days 2018 data set.
It is clear that temperature variations may affect the apparatus in many different ways.
They change the geometry of the gyroscope, but also deform the monument, generating spurious rotation. It is not straightforward to separate one effect from the other.
Aim the present analysis is then to understand the effective role of as many as possible thermal contribution to the instrument signals.
June 2018 data set has a nice clean temperature behaviour with the maximum excursion limited to less than 0.03~$^{\circ}$C. 
Then we will use these data to investigate residual contribution of the temperature in $\Omega_{local}$.

As clearly visible in Fig.\ref{fig:Omega_T}, $\Omega_{local}$ on quite long time scales follows the temperature behavior.
This is a clear indication of a possible linear relation between the two variables that can be interpreted as a consequence of the bare instrument expansion due to thermal drifts.
From Eq. (\ref{eq:general}) we see that the SF depends linearly on the ring side length. We can, then, assume that for
long term operations, the relative effect of a temperature variation $\Delta T$ on the Sagnac frequency is given by $\rho\cdot\Delta T$, where $\rho$ is the thermal expansion coefficient.
In our case, the expansion coefficient is the granite one $\rho_{g} = 6.5\cdot 10^{-6}/\,^{\circ}$C and the average Sagnac frequency is $f_s=280$ Hz. 
The expected effect is of the order of $f_s \cdot\rho\cdot\Delta T$~Hz. A linear regression carried on the 2018 data set, after separating local and global signals, 
gives a temperature coefficient of $0.7\cdot 10^{-3}$~Hz/$^{\circ}$C, to be compared to the expected value of 1.8$\cdot 10^{-3}$ Hz/$^{\circ}$C.
	 
\begin{figure*}
    \centering
    \includegraphics[width=\textwidth] {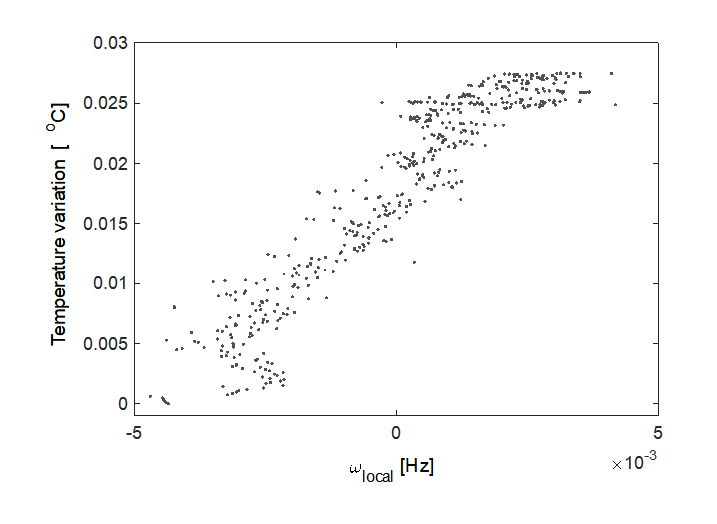} 
    \caption{Relationship between $\frac{\omega_{local}}{2\pi}$ and temperature variations.}
    \label{fig:tempjune}
\end{figure*}

Another approach is to scatter plot the two variables, $\Omega_{local}$ and $T$  in a region where the temperature varies linearly. 
This can be done selecting the data relative to the upward slope of the first $\sim3$ days. 
Fig. \ref{fig:tempjune} shows in a scatter plot $\Omega_{local}$ vs. T, the linear relation in this case is of the order of 0.45~Hz/$^{\circ}$C, as obtained by a linear fit on the plotted data.
This may indicate that temperature induces perturbations in the apparatus much larger than the geometrical scale factor change thus confirming that temperature effects are multifaceted.
Eventually, we note that the same approach is not worthy for the remaining of the data of the 2018 run.
As a matter of fact, faster change in the temperature suppresses the relative scattering of $\omega_{local}$ but, 
once the temperature get more quite, the scatter around the mean relatively increases.

\subsection{Interplay between temperature and tilts}
\label{mixing}

To further understand the role of the temperature we have investigated the relation between temperature and tilt signals.
In this case, as it is evident comparing the time series in Figs. \ref{fig:Omega_T} and \ref{fig:Tilt_s} there is no evidence of a simple relation. 
However, it is plausible that temperature changes induce some spurious rotation of the apparatus due to any anisotropy of the mechanical structure and/or of the underneath concrete monument.
To appreciate whether or not an effect arises in the tilts we assumed a third order polynomial in the temperature variable as the driving for tilts. 
Consequently, we have, at first, calculated the maximum tilt direction for each time point, 
then fitted these tilt values with a polynomial fit of the third order leaving the temperature as explanatory variable.

These procedure has a twofold aim. On one hand, modelling the contribution to tilts coming from the temperature due to instrumental deformation allows to better use tilt-meter data in calculating the final instrument accuracy. On the other hand, the standard deviation of the residuals give an evaluation of the orientation stability of the cave itself.
Standard deviations of the residuals, indeed, gives the fraction of inclination not related to temperature in this model.

We found a residual standard deviation of 0.726~$\mu$rad for the 2018 30 days series 
(third order fit r-squared=0.707 with 0.01~$^{\circ}$C thermal stability, and 
$\pm2.5\ \mu$rad inclination range and STD 1.3~$\ \mu$rad) while it goes down to 0.25$\ \mu$rad in the case of 103 days 2020 (third order fit r-squared=0.891 
with 0.1~$^{\circ}$C thermal stability, $\pm 1.5\, \mu$rad range, and STD 0.73\ $\mu$rad). 
From this result we can infer that the orientation of the underneath bedrock is stable at the level of microradian.

The relation between  tilts and temperature in some portion of data has been estimated of the order of 500~$\mu$rad/$^{\circ}$C.
In GINGER, the RLG at maximum signal requires a long term stability of the monument of the order 1~$\mu$rad  \cite{angela2017}. 
If we assume in GINGER a long term temperature stability of 0.01~$^{\circ}$C, we need to improve at least a factor 5 the orientation stability of the monument. 
We expect that in GINGERINO the changes of orientation are mostly related to the non uniformity of the basement which is connected to the bedrock through a reinforced concrete block, whose homogeneity was not cured.

We have also investigated possible effects of the temperature derivative $dT/dt$. It could deform the shape of the HL mechanical structure, 
producing a spurious rotation of the ring mirrors. This derivative, however, is a very small quantity, at the limit of the measurement noise level. 
To be sure that, at the present sensitivity, GINGERINO does not see any effect of this term we looked at first 10 days data of June 2018, 
showing larger temperature drift being acquired soon after a closing of the box. 
There we had for $dT/dt$ a maximum value of the order of $10^{-7}$ $^{\circ}$C/s but we did not find a clear relation between the gyroscope signal and $dT/dt$.
In any case, we expect that this effect is very small, and can be further reduced by an active control of the ring geometry 
and by improving the isolation of the apparatus from external perturbation.

Typical long term (some weeks) temperature fluctuations are of the order of $10^{-2}$~$^{\circ}$C, but in the 2020 long run
we observed only $\sim 0.1$~$^{\circ}$C stability, ten times worse. 
This is probably due to deterioration of the protection box, which we will repair as soon as possible.
\footnote{Access to the experiment is presently limited due CoVid pandemic.
The whole GINGERINO experiment is enclosed in a box, to isolate from the cave. The box is wormed up by infrared lamps in order to keep the relative humidity below $60\%$}

\subsection{Coupling between the rotation of the HL RLG and the inclination of the monument}
\label{inclination}

\begin{figure*}
    \centering
     \includegraphics[width=\textwidth] {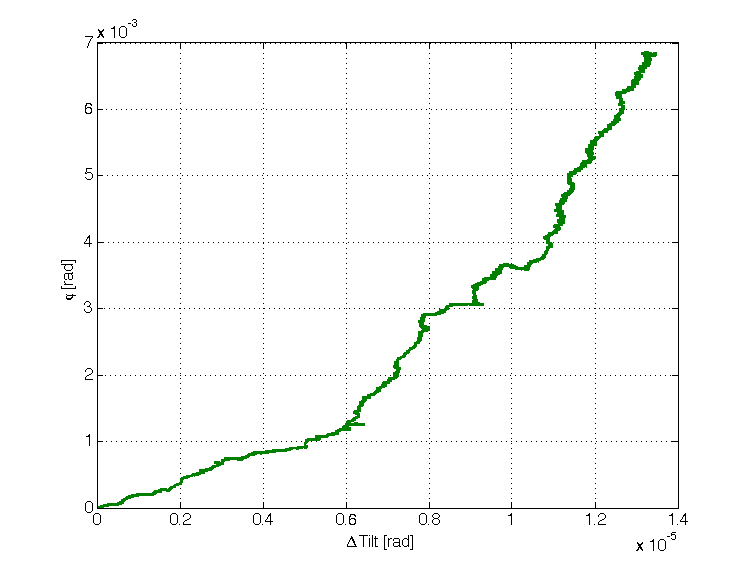}  
    \caption{Scatter plot of the absolute value of the change of inclination $\Delta_{Tilt}$ versus the phase variation $\phi$ associated to the tilt.}
    \label{fig:ROT}
\end{figure*}

GINGERINO is based on one of the first mechanical model for heterolithic RLG, usually called GEOSENSOR, developed for application in seismology. 
The mirrors can be easily aligned using mechanical levers in air, which act on the mirror boxes. This smart and convenient solution 
has the drawback that all the mechanical parts form a continuous object, and basically very small rotation of one part can effectively make the cavity rotate.
This means that any tilt and/or mechanical stress due to temperature may induce a spurious, tiny and slow, rotation of the optical cavity. 
Under this assumption we expect to observe a phase variation $\phi$ associated to the tilt. 
This variation can be reconstructed integrating in time $\omega_{local}$, at this purpose we remind that the data are acquired at 600s rate, and before integration, 
interpolated in order to fill the gaps due to the missed points, associated with mode jumps and split mode operation.
In Fig. \ref{fig:ROT} we report the change of the absolute value of inclination $\Delta Tilt$ versus the phase $\phi$ for the 2020 data, the longer data set. The plot shows a clear correlation between the two variable indicating that whenever the ring structure tilts we see a rotation of the optical cavity.

Similar behaviour is found using the other data sets.
To have a quantitative estimation of the connection between tilt and rotation we have run a linear fit on the plotted data. 
The fit indicates that there is a linear relationship of the phase with the inclination of the monument, of the order of 
$550\pm5$~rad per rad of inclination (r-squared = 0.93).
This connection between tilts and instrumental rotation will be reduced by the new mechanical design we are developing for GINGER. In GINGER each mirror will be uncoupled from the rest of the mechanics.
Active controls, absent in GINGERINO, on mirrors position and tilt should be designed in order to avoid couplings to the entire structure. Moreover, in order to control and eventually subtract this contribution, it is possible to develop system to monitor the position of each mirror with respect to the granite support.

\section{Conclusions}
\label{Sect:Concl}

GINGERINO is the first underground HL RLG operative in a continuous basis, with sensitivity better than prad/s. RLG signal is the sum of a global contribution coming from the Earth rotation and a local one that contains geophysical signals and local and instrument disturbances. 
We have investigated the fingerprints of environmental parameters, such as pressure, temperature and local tilts, in the local contribution to the Sagnac frequency as measured by GINGERINO.  
Our study proves the very close relation between temperature and Sagnac signal. 
It is not only given by the thermal expansion of the granite support, which changes the perimeter of the cavity and accordingly the geometrical scale factor of the Sagnac gyroscope. 
However, the coupling of temperature variations seems to be more complicate affecting mainly the support structure and the HL mechanical structure of GINGERINO.
By looking at the local tilts, measured by a two-axis tilt-meter attached to the granite support, 
we found evidence of a coupling between this degrees of freedom with the temperature variations.
It most probably comes from the reinforced concrete interface between the granite and the underneath bedrock that is not homogeneous and so its orientation changes with temperature variation. 
The study of the local disturbances $\omega_{local}$ shows that the RLG cavity rotates when the monument tilts, this is associated with the HL mechanical design of GINGERINO, in which the mirrors are not fixed to the monument.
Cross checking the temperature behaviour and the tilt-meter signal we also proved that the residual orientation stability at the level of $\mu$rad at LNGS is suited for the construction of GINGER the RLG array able to measure General Relativity effects. 
To reach the required sensitivity and accuracy improvement in the mechanics, especially for the structure underneath the ring, are necessary. The present study indicate the path to follow toward GINGER: increase the instrument isolation; improve the holding structure homogeneity; reduce the coupling between tilt and cavity rotation. All of these have a feasible solution.

\begin{acknowledgement}
We thank the Gran Sasso staff in support of the experiments, particularly Stefano Gazzana, Nazzareno Taborgna and Stefano Stalio .
We are thankful for technical assistance to   Alessio Sardelli and Alessandro Soldani of INFN Sezione di Pisa and Francesco Francesconi of Dipartimento di Fisica.
A special thank to Gaetano De Luca of Istituto Nazionale di Geofisica e Vulcanologia for regularly checking Gingerino operation status.
\end{acknowledgement}

\bibliographystyle{unsrt}
\bibliography{library}


\end{document}